# Hadamard Product Decomposition and Mutually Exclusive Matrices on Network Structure and Utilization

Michael Ybañez[1], Kardi Teknomo[2], Proceso Fernandez[3]

Department of Information Systems and Computer Science, Ateneo de Manila University
Quezon City, Philippines
[1]mybanez@ateneo.edu
[2]teknomo@gmail.com
[3]pfernandez@ateneo.edu

*Abstract*— **Graphs are very important mathematical structures used in many applications, one of which is transportation science. When dealing with transportation networks, one deals not only with the network structure, but also with information related to the utilization of the elements of the network, which can be shown using flow and origin-destination matrices. This paper extends an algebraic model used to relate all these components by deriving additional relationships and constructing a more structured understanding of the model. Specifically, the paper introduces the concept of mutually exclusive matrices, and shows their effect when decomposing the components of a Hadamard product on matrices.**

## I. INTRODUCTION

Graphs in computer science have many different applications. In the specific context of traffic and transportation science, graphs are used to represent transportation pathways and are used extensively for urban planning schemes. As additional information used in tandem with transportation network, trajectory data from pedestrians and other elements of traffic such as cars, motorcycles, and other vehicles are gathered through several tracking methods which usually involve the usage of GPS sensors. Because all of these concepts are related, it is useful to see if it is possible to discover close relationships between the network data and the trajectory data, in order to arrive at better methods for deriving one from the other.

The previous study of Teknomo and Fernandez [1] divides the network analysis into two components – network structure and network usage. Network structure corresponds to the static part of the network (such as the road network), while the network usage corresponds to the dynamic component (e.g. vehicular movements). Several matrices were described in order to capture some important concepts in each of these components. The described matrices were further analysed in three levels – set level, count level and binarization level.

Trajectories are useful for aggregating information into traffic flow, which can then be used to obtain the origin destination (OD) matrix. In fact, [1] linked the OD matrix, flow, and trajectories into a unified mathematical model.

This paper extends that model to better strengthen these relationships, and shows additional interesting conclusions regarding some of the elements. Specifically, we define mutual exclusivity for matrices, and explore pairs of matrices that satisfy this property. The additional relationships among matrices that were surfaced are then used to decompose further the analysis of the interaction between network structure and usage.

## II. PRELIMINARIES

In this section, we define important terms and values in this paper, and present algorithms for acquiring some of these values as described in [1].

A trajectory is the path or route taken by a moving agent or traffic element within a specified observation period from $t_1$ to $t_2$, where $t_1 < t_2$. This path is denoted by a sequence of points that the agent travels through. For the purposes of this paper, we assume that agents do not visit the same point more than once; thus, the trajectories have no cycles.

A trajectory in a practical sense can be mapped to a latitude and longitude, but for the sake of simplification, it is helpful to map trajectories to network graphs. When only the relative order of visited points (or nodes) is recorded, and the exact time of visits is discarded. These trajectories are called ordinal graph trajectories. With these trajectories, we can now deal with several pertinent network-related structures, defined below.

Given a network with nodes, the representation of the network is given by an adjacency matrix $\mathbf{A}$, where each element in the matrix has a binary (0,1) value to represent the absence or presence of an edge or link between pairs of the nodes of the network. The distance (path) matrix $\mathbf{P}$ is another matrix whose elements contain the length of the shortest path between the corresponding nodes of the network. These matrices are standard in graph theory literature. The adjacency matrix is given for any graph, while the distance matrix can be easily computed using the Floyd-Warshall algorithm in $\mathcal{O}(k^2)$ time. As a note, this paper deals with directed graphs, so the adjacency matrix is not necessarily symmetric.

We also define a third matrix: the external matrix **E**. We define it to be the matrix computed from the difference between **P** and **A**. This operation is a simple element-wise subtraction and thus **E** can be also computed in $\mathcal{O}(k^2)$ time.

The above three matrices are considered to be static, as they do not generally change during observation. They are matrices that represent structural properties of the network. In many cases, we may simply be interested in a binarized form of the matrices. The binarized forms of **P** and **E** are represented by $\widehat{\mathbf{P}}$ and $\widehat{\mathbf{E}}$, respectively. The binary matrix $\widehat{\mathbf{P}}$ is defined as:

$$\widehat{\mathbf{p}}_{i,j} = \begin{cases} 1 \text{ if } 0 < \widehat{\mathbf{p}}_{i,j} < \infty \\ 0 \text{ otherwise} \end{cases}$$

The inverted breve operator on the matrices represents binarization of those matrices. Note that $\infty$ is the symbol given for when no path exists between two nodes, when computed by the Floyd-Warshall algorithm. The binarized form $\widehat{\mathbf{E}}$ of the external matrix is defined similarly to the above definition for $\widehat{\mathbf{P}}$.

We can now define several additional matrices related not to the structure of the network, but to the utilization of this network by the trajectories. The flow matrix **F** is the matrix whose elements contain the number of trajectories (flows) going from some source node to some sink node directly (i.e., using the direct link or edge between the source and sink.) Note that any trajectory using the edge $(i, j)$ contributes to the flow matrix element $\mathbf{f}_{i,j}$. The OD matrix **D** is the origin-destination matrix, whose elements contain the number of trajectories from a source to a sink using any possible path in the network. In contrast to the flow matrix, the OD matrix cares only about how many trajectories go from source to sink nodes, but disregards their choice of path when in the presence of multiple possible paths to take. Traditionally, OD matrices only deal with specific source and sink nodes (chosen for their importance in a network), but this model uses a generalized OD matrix that tracks information on all pairs of nodes in the network.

The indirect flow matrix **L** is similar to **F**, except it counts only those trajectories that explicitly do not use the direct link between the given pair of nodes. Two more matrices related to **L** are defined: **T** is the alternative route flow matrix which counts indirect flows between nodes where a direct link actually exists but is not chosen by the moving element (trajectory), and $\mathbf{T}^C$ is the substitute route flow matrix which counts indirect flows between nodes where no direct link or edge exists between these nodes.

The above five matrices **F**, **D**, **L**, **T**, and $\mathbf{T}^C$, deal with the utilization of the network. They also have binary forms defined similarly, and are denoted by $\widehat{\mathbf{F}}$, $\widehat{\mathbf{D}}$, $\widehat{\mathbf{L}}$, $\widehat{\mathbf{T}}$, and $\widehat{\mathbf{T}^C}$ respectively.

### III. RELATED LITERATURE

In the context of transportation planning, there are many different types of analysis available and many different scientific problems to tackle. One example of these problems is the traffic assignment problem, which attempts to predict the traffic within a given network. From observed data, an origin-destination (OD) matrix is derived, and its elements refer to the expected number of traffic elements (trajectories) going from some origin node to some destination node. On each of the links, we can also create a measure of the traffic or flow in that link, which we encapsulate in a flow matrix. The traffic assignment problem deals with predicting the flow within the links of the network given the OD matrix information. This type of prediction is very important for many fields, such as urban planning in the context of road and street networks.

There are many studies in the field of transportation science that deal with methods of acquiring these matrices and other related information from real-life data. One important piece of information is trajectory data. Some methods use GPS sensors to capture precise position data and process it automatically to generate trajectory information [2]. Other methods use speed data instead for determining trajectories of vehicles [3]. Other methods involve image processing to track vehicles, and even to classify them into different types, such as trucks and motorcycles [4]. After data is gathered through different methods, software packages created [5] may be used to analyze the data automatically.

In many cases, traffic data that is gathered is essential to creating good estimates and predictions on transportation networks. Much research has been devoted to estimating origin-destination matrices with traffic counts as input, and there are already several classical solutions to this problem, and variants thereof [6] [7]. Many methods attempt to relieve humans of costly data-gathering methods. Instead of using surveys, observed link flows [8] or traffic counts on intersections [9] can be used instead. Many different methods also have different considerations, or may attempt to do different variations of the problem. Some methods do OD estimation that considers multiple-vehicle data [10], while others attempt to do dynamic estimation [11].

### IV. PREVIOUS MODEL

The previous model [1] showed the relationships between the matrices **A**, **E**, **P**, **F**, **D**, **L**, **T**, and $\mathbf{T}^C$, for different forms: matrix, binarized matrix, and matrix-set. We can divide them into *structural matrices* and *utilization matrices* (also called *usage matrices*). The structural matrices are the following: **A**, the adjacency matrix; **E**, the external matrix; and **P**, the distance matrix. The utilization matrices are the following: **D**, the generalized OD matrix; **F**, the direct flow matrix; **L**, the indirect flow matrix; **T**, the alternative route flow matrix; and $\mathbf{T}^C$, the substitute route flow matrix.

The previous model derived two sets of equations and inequalities; one that is valid for all instances, and one that is valid only for fully utilized networks. (Fully utilized networks are networks for which every link in the graph is used by at least one trajectory.) Table 1 below shows a summary of these results.

We take note of several notational points regarding the results of the previous model: First, the $\widetilde{\mathbf{X}}$ notation for some matrix **X** refers to the matrix-set-level analysis for that matrix; however, this is beyond the scope of this paper. Second, the ∘ operator refers to the Hadamard product of two matrices, which is an element-wise multiplication of the elements of the matrix operands[1]. Third, the term *fully utilized* refers to a network for which each of the edges is traversed by at least one trajectory. The results of this paper do not make any distinction between fully utilized and non-fully utilized networks, and are thus generalized for all types of networks.

TABLE I
SUMMARY OF THE MAIN RESULTS OF [1]

| Valid for all instances | Valid only for instances involving fully utilized networks |
|---|---|
| $\widehat{\mathbf{E}} = \widehat{\mathbf{P}} \circ \widehat{\mathbf{E}}$ and $\mathbf{A} = \widehat{\mathbf{P}} \circ \mathbf{A}$ | |
| $\widehat{\mathbf{F}} \leq \mathbf{A}$ and $\mathbf{F} = \mathbf{A} \circ \mathbf{F}$ | $\widehat{\mathbf{F}} = \mathbf{A}$ |
| $\widehat{\mathbf{D}} \leq \widehat{\mathbf{P}}$ and $\mathbf{D} = \widehat{\mathbf{P}} \circ \mathbf{D}$ | $\widehat{\mathbf{D}} = \widehat{\mathbf{P}}$ |
| $\widetilde{\mathbf{F}} \subseteq \widetilde{\mathbf{D}}$, $\mathbf{F} \leq \mathbf{D}$ and $\widehat{\mathbf{F}} \leq \widehat{\mathbf{D}}$ | |
| $\widetilde{\mathbf{T}} = \mathbf{A} \circ \widetilde{\mathbf{L}}$, $\mathbf{T} = \mathbf{A} \circ \mathbf{L}$ and $\widehat{\mathbf{T}} = \mathbf{A} \circ \widehat{\mathbf{L}}$ | |
| $\mathbf{F} + \mathbf{T} = \mathbf{A} \circ \mathbf{D} = \mathbf{D} - \mathbf{T}^C$ | |
| $\mathbf{T} = \mathbf{A} \circ \mathbf{D} - \mathbf{F}$ and $\mathbf{T}^C = \widehat{\mathbf{E}} \circ \mathbf{D}$ | |
| $\mathbf{L} = \mathbf{T} + \widehat{\mathbf{E}} \circ \mathbf{D}$ $\Leftrightarrow \mathbf{A} \circ \mathbf{D} = \mathbf{D} - \widehat{\mathbf{E}} \circ \mathbf{D}$ $\Leftrightarrow \mathbf{D} = \widehat{\mathbf{P}} \circ \mathbf{D}$ $\Leftrightarrow \mathbf{F} = \mathbf{A} \circ \mathbf{D} - \mathbf{T}$ | |

V. EXTENSIONS TO THE MODEL

This paper shows some additional relationships between the different matrices in the model. These relationships are grouped into the following meaningful categories.
  a. Mutually exclusive matrices
  b. Substitute route flow-related equations
  c. Alternative route flow-related equations
  d. Total indirect flow-related equations
  e. Other equations
  f. Exceptions and wrong equations
We shall now describe each of the categories.

*A. Mutually Exclusive Matrices*

In this study, we introduce the concept of mutually exclusive matrices. Two *n* x *m* matrices **X** and **Y** are said to be mutually exclusive if and only if the following holds:

$$\mathbf{x}_{i,j} \neq 0 \Rightarrow \mathbf{y}_{i,j} = 0 \ \forall 1 \leq i \leq n, 1 \leq j \leq m$$

---
[1] We must be careful not to try dividing the equations involving Hadamard products. For example, the equation $\mathbf{F} = \mathbf{A} \circ \mathbf{F}$ does not indicate that in all cases $\mathbf{A} = \mathbf{1}$ (the all-ones matrix); in fact, entrywise division is problematic if the divisor contains zero entries, which in the field of transportation networks is almost always true.

That is, the two matrices do not have corresponding nonzero elements. Consequently, the Hadamard product of the two matrices is a zero matrix.

$$\mathbf{X} \circ \mathbf{Y} = 0$$

Mutually exclusive matrices capture the idea that corresponding elements of two matrices cannot co-exist.

*1) Adjacency matrix and external matrix:* The first examples of mutually exclusives matrices we can show are the adjacency matrix and the binarized external matrix. We use the binarized form because it disregards any length, as well as eliminates any problems with the ∞ symbol, such as when we attempt to multiply ∞ with 0, which is undefined. We note that if the elements of a matrix are only zeros or ones, then multiplying a matrix with itself will result in the same matrix. We also note that the adjacency matrix is binarized, by definition. Our first result may be expressed in the following equation.

$$\mathbf{A} \circ \widehat{\mathbf{E}} = \mathbf{0} \quad (0)$$

Equation (0) has two operands that describe the structure of the network. Recall that $\widehat{\mathbf{E}}$ is given by $\widehat{\mathbf{E}} = \widehat{\mathbf{P}} - \mathbf{A}$, where $\widehat{\mathbf{P}}$ is the binarized path matrix. This means that $\widehat{\mathbf{P}}$ simply represents whether nodes are reachable from other nodes, and not the minimum number of edges necessary for such a traversal. The previous study also proved that $\mathbf{A} = \widehat{\mathbf{P}} \circ \mathbf{A}$. Using these, Equation (0) can now be derived as follows:

$$\mathbf{A} \circ \widehat{\mathbf{E}} = (\widehat{\mathbf{P}} \circ \mathbf{A})(\widehat{\mathbf{P}} - \mathbf{A})$$
$$= \widehat{\mathbf{P}}\widehat{\mathbf{P}}\mathbf{A} - \widehat{\mathbf{P}}\mathbf{A}\mathbf{A}$$
$$= \widehat{\mathbf{P}}\mathbf{A} - \widehat{\mathbf{P}}\mathbf{A}$$
$$= \mathbf{0}$$

Thus, the adjacency matrix and the binarized external matrix are mutually exclusive.

*2) Structure vs. Usage*

$$\mathbf{A} \circ \mathbf{T}^C = \mathbf{0} \quad (0)$$
$$\mathbf{A} \circ \widehat{\mathbf{T}}^C = \mathbf{0} \quad (0)$$
$$\mathbf{F} \circ \widehat{\mathbf{E}} = \mathbf{0} \quad (0)$$
$$\widehat{\mathbf{F}} \circ \widehat{\mathbf{E}} = \mathbf{0} \quad (0)$$
$$\mathbf{T} \circ \widehat{\mathbf{E}} = \mathbf{0} \quad (0)$$
$$\widehat{\mathbf{T}} \circ \widehat{\mathbf{E}} = \mathbf{0} \quad (0)$$

The above six equations show that there are matrices for which the Hadamard product is the zero matrix. The Hadamard product is the result of element-wise multiplication for 2 matrices having the same dimensions. We will show the proofs for the above equations, and also describe their significance.

Equation (0) can be proven as follows: For any two nodes *i*, *j*, either there is an edge (*i*, *j*) between them, or there is no edge. In the first case, $a_{i,j} = 1$, but by the definition of $\mathbf{T}^C$, no substitute route flow can exist between the nodes if a link exists between the two nodes. Therefore, $t^C_{i,j} = 0$ in this case, and the product $a_{i,j} \cdot t^C_{i,j} = 0$. In the second case, $a_{i,j} = 0$ because there is no edge between the two nodes; thus, we also get a product of 0.

For equation (0), the proof is done similarly, except that $\mathbf{T}^C$ will always have either 0 or 1 for each of its elements, and the corresponding elements for $\mathbf{A}$ will have the other value. This implies that the product will always be the zero matrix.

The proof of equation (0) is similar to the proof of the above two equations. The matrices $\widehat{\mathbf{E}}$ and $\mathbf{F}$ can be shown to be also mutually exclusive: $\mathbf{f}_{i,j} = 0$ when $\mathbf{a}_{i,j} = 0$ (because a flow on some link $(i, j)$ cannot exist if the link itself does not actually exist), so multiplying $\mathbf{f}_{i,j} = 0$ with either zero or one is still zero. In the second case, if $\mathbf{a}_{i,j} = 1$, then there may be trajectories that exist that utilize the edge $(i, j)$, so $\mathbf{f}_{i,j} \geq 0$. However, $\mathbf{e}_{i,j} = 0$ if $\mathbf{a}_{i,j} = 1$, so the product is still zero.

The proofs for equations (0), (0), and (0) are similar to the proofs outlined above and will be omitted here. As a note, the alternative route matrix $\mathbf{T}$ is similar to $\mathbf{F}$ in that trajectories may contribute to the number of alternative routes taken from some node $i$ to some node $j$ only if a direct link $(i, j)$ exists.

[1] described how certain matrices ($\mathbf{A}, \mathbf{P}, \mathbf{E}$) represent some aspects of the structure of the graph, and how others (such as $\mathbf{T}$ and $\mathbf{T}^C$) describe the usage of the network by trajectories. We notice that the above six equations multiplied a structural matrix by a usage or utilization matrix. Although it is not always the case that multiplying a structural matrix by a usage matrix will result in the zero matrix, the above equations show tighter relationships between some pairs of structural matrices and utilization matrices.

First, the adjacency matrix $\mathbf{A}$ and the external matrix $\widehat{\mathbf{E}}$ are both structural matrices that are mutually exclusive, as shown in equation (0). From the definitions of $\mathbf{F}$ and $\mathbf{T}$, it is apparent that direct flows and alternative route flows can only exist in the presence of direct links. Thus, these "derivatives" of $\mathbf{A}$ are also mutually exclusive with $\mathbf{E}$, because when they are multiplied by the external matrix $\mathbf{E}$, will always result in the zero matrix (see equations (0) through (0) above).

Equations (0) and (0), however, suggest that the (binarized) external matrix $\widehat{\mathbf{E}}$ is closely linked to the substitute route flow matrix $\mathbf{T}^C$. Though the original model suggested that $\widehat{\mathbf{E}}$ is associated with the indirect flow matrix $\mathbf{L}$, the above relationships would suggest that a tighter association lies elsewhere. Based on the properties of $\widehat{\mathbf{E}} = \widehat{\mathbf{P}} - \mathbf{A}$ in relation to $\widehat{\mathbf{P}}$ and $\mathbf{A}$, specifically, that $\widehat{\mathbf{e}}_{i,j} = 1$ if and only if $\widehat{\mathbf{p}}_{i,j} = 1$ but $\mathbf{a}_{i,j} = 0$, then it appears that $\widehat{\mathbf{E}}$ is the structural matrix that directly corresponds to the usage matrix $\mathbf{T}^C$: by definition of a substitute route flow, we see that $\mathbf{t}^C_{i,j} \geq 1$ if and only if there is a path from node $i$ to node $j$ (meaning $\widehat{\mathbf{p}}_{i,j} = 1$) but there is no direct link between them ($\mathbf{a}_{i,j} = 0$). This means that $\mathbf{t}^C_{i,j} \geq 1$ only if $\widehat{\mathbf{e}}_{i,j} = 1$. An informal way of stating this relationship is that $\widehat{\mathbf{E}}$ represents the possibility of having substitute route flows ($\mathbf{T}^C$): if $\widehat{\mathbf{e}}_{i,j} = 0$, no substitute route flows from $i$ to $j$ can exist.

We can then see how equations (0) and (0) make sense: when we multiply $\mathbf{A}$ by a "derivative" of $\widehat{\mathbf{E}}$, specifically $\mathbf{T}^C$, then we produce the zero matrix.

We observe that in addition to $\mathbf{A}$ and $\widehat{\mathbf{E}}$ being mutually exclusive, we can get more pairs of mutually exclusive matrices by getting a usage matrix that is the derivative of either $\mathbf{A}$ or $\widehat{\mathbf{E}}$, as seen in the above equations. This usage matrix may or may not be binarized, but the relation still holds.

*3) Usage vs. Usage*

$$\mathbf{F} \circ \mathbf{T}^C = \mathbf{0} \qquad (0)$$
$$\mathbf{F} \circ \widehat{\mathbf{T}}^C = \mathbf{0} \qquad (0)$$
$$\widehat{\mathbf{F}} \circ \mathbf{T}^C = \mathbf{0} \qquad (0)$$
$$\widehat{\mathbf{F}} \circ \widehat{\mathbf{T}}^C = \mathbf{0} \qquad (0)$$

We now investigate the case where both the matrices involved are usage or utilization matrices which are originated from $\mathbf{A}$ and $\widehat{\mathbf{E}}$. We select $\mathbf{F}$ as a usage matrix derived from $\mathbf{A}$: we notice that by definition, a flow $\mathbf{f}_{i,j}$ may only exist if $\mathbf{a}_{i,j} = 1$, that is, an edge $(i, j)$ exists. We also select $\mathbf{T}^C$ as a usage matrix for $\widehat{\mathbf{E}}$ (see the previous subsection for more details on this relationship). We show a proof for equation (0), and skip the proofs for the other equations as they simply involve the binarized forms of the matrices, and thus have similar proofs.

[1] proved that $\mathbf{F} = \mathbf{A} \circ \mathbf{F}$ and $\mathbf{T}^C = \widehat{\mathbf{E}} \circ \mathbf{D}$. Our proof for equation (0) then goes as follows:

$$\mathbf{F} \circ \mathbf{T}^C = (\mathbf{A} \circ \mathbf{F}) \circ (\widehat{\mathbf{E}} \circ \mathbf{D})$$
$$= (\mathbf{A} \circ \widehat{\mathbf{E}}) \circ \mathbf{F} \circ \mathbf{D}$$
$$= \mathbf{0} \circ \mathbf{F} \circ \mathbf{D}$$
$$= \mathbf{0}$$

The above proof is a direct result of the mutual exclusivity of $\mathbf{A}$ and $\widehat{\mathbf{E}}$.

As an alternative proof, we can also prove this through an element-wise derivation: Given two nodes $i$ and $j$, either the edge $(i, j)$ exists, or it does not. If it does, then $\mathbf{t}^C_{i,j} = 0$ because no substitute route flows can exist when a direct edge exists, as by definition. The value of $\mathbf{f}_{i,j}$ is then irrelevant, as the product will still be zero. In the second case, since there is no link between $i$ and $j$, then $\mathbf{f}_{i,j} = 0$, and the product will remain zero. A similar proof can be done for the other equations involving the binarized forms of $\mathbf{F}$ and $\mathbf{T}^C$.

We thus see that $\mathbf{F}$ and $\mathbf{T}^C$, both usage matrices, are mutually exclusive matrices.

Another set of exclusive matrix pairs may be derived from the definition-based exclusivity property of the alternative route and substitute matrices $\mathbf{T}$ and $\mathbf{T}^C$.

$$\mathbf{T} \circ \mathbf{T}^C = \mathbf{0} \qquad (0)$$

$$\widehat{\mathbf{T}} \circ \widehat{\mathbf{T}}^C = \mathbf{0} \tag{0}$$

$$\widehat{\mathbf{T}} \circ \mathbf{T}^C = \mathbf{0} \tag{0}$$

$$\mathbf{T} \circ \widehat{\mathbf{T}}^C = \mathbf{0} \tag{0}$$

The above equations can be readily shown to be true using the definitions of alternative route flows and substitute route flows: the first can only exist when a direct edge exists, and the second only when a direct edge does not exist.

### B. Substitute route flow-related equations

We will now show equations related to the substitute route flow matrix $\mathbf{T}^C$.

$$\widehat{\mathbf{D}} \circ \mathbf{T}^C = \mathbf{T}^C \tag{0}$$

$$\widehat{\mathbf{L}} \circ \mathbf{T}^C = \mathbf{T}^C \tag{0}$$

$$\widehat{\mathbf{E}} \circ \mathbf{T}^C = \mathbf{T}^C \tag{0}$$

$$\widehat{\mathbf{T}}^C \circ \mathbf{T}^C = \mathbf{T}^C \tag{0}$$

The above four equations can be easily shown to be true. Whenever $\mathbf{t}_{i,j}^C \geq 1$, there is at least one trajectory that goes from node $i$ to node $j$ indirectly; therefore, both $\mathbf{d}_{i,j} \geq 1$ and $\mathbf{l}_{i,j} \geq 1$, and $\widehat{\mathbf{d}}_{i,j} = \widehat{\mathbf{l}}_{i,j} = 1$. (The case $\mathbf{t}_{i,j}^C = 0$ is trivial because multiplication by zero results in zero, which preserves the equality.) This proves equations (0) and (0). Because $\mathbf{T}^C$ also implies the absence of a direct link between $i$ and $j$, then $\mathbf{e}_{i,j} \geq 1$ and $\widehat{\mathbf{e}}_{i,j} = 1$, thus proving equation (0). Finally, $\mathbf{t}_{i,j}^C = 1$ whenever $\mathbf{t}_{i,j}^C \geq 1$, and a multiplication by 1 preserves the equality, proving equation (0). Note that the above equations multiply $\mathbf{T}^C$ with a binarized matrix, and all of them except for $\widehat{\mathbf{E}}$ are usage matrices.

We now investigate cases where we instead multiply $\mathbf{T}^C$ with other usage matrices.

$$\mathbf{D} \circ \mathbf{T}^C = \mathbf{T}^C \tag{0}$$

$$\mathbf{L} \circ \mathbf{T}^C = \mathbf{T}^C \tag{0}$$

$$\widehat{\mathbf{D}} \circ \mathbf{T}^C = \mathbf{T}^C \tag{0}$$

$$\widehat{\mathbf{L}} \circ \mathbf{T}^C = \mathbf{T}^C \tag{0}$$

We note that $\mathbf{L}$ is the indirect route flow matrix, which may be decomposed into two mutually exclusive matrices, $\mathbf{T}$ and $\mathbf{T}^C$, using the formula $\mathbf{L} = \mathbf{T} + \mathbf{T}^C$. (This is clearly shown by the fact that there is either a direct link between any two nodes, or there is none.) We observe that whenever a substitute route flow exists, i.e., $\mathbf{t}_{i,j}^C = 1$, then $\mathbf{t}_{i,j} = 0$ by definition. Because $\mathbf{L} = \mathbf{T} + \mathbf{T}^C$, then if a substitute route flow exists, $\mathbf{l}_{i,j} = \mathbf{t}_{i,j}^C$, and the product in equation (0) holds. Equation (0) can be analyzed similarly: $\mathbf{d}_{i,j}$ consists of direct flows $\mathbf{f}_{i,j}$, alternative route flows $\mathbf{t}_{i,j}$, and substitute route flows $\mathbf{t}_{i,j}^C$. However, if substitute route flows exist for some pair of nodes $i$ and $j$, then the edge $(i, j)$ does not exist, and thus no direct flows or alternative route flows exist. This means that if $\mathbf{t}_{i,j}^C = 1$, then $\mathbf{d}_{i,j}$ pertains only to the number of substitute route flows, and the product holds. The proofs for equations (0) and (0) are similar and are thus omitted.

$$\widehat{\mathbf{E}} \circ \mathbf{T}^C = \mathbf{T}^C \tag{0}$$

Equation (0) shows a relationship between $\widehat{\mathbf{E}}$ and $\mathbf{T}^C$. When $\mathbf{t}_{i,j}^C = 1$, this means that a substitute route flow exists; this implies that there is no edge $(i, j)$ yet a path from $i$ to $j$ still exists. Also, by definition of the external matrix, $\widehat{\mathbf{e}}_{i,j} = \widehat{\mathbf{p}}_{i,j} - \mathbf{a}_{i,j} = 1$ if and only if $\widehat{\mathbf{p}}_{i,j} = 1$ and $\mathbf{a}_{i,j} = 0$. This means that if a substitute route flow exists from node $i$ to node $j$, then $\widehat{\mathbf{e}}_{i,j} = 1$, and the product holds for this case. In the case where $\mathbf{t}_{i,j}^C = 0$, no substitute route flow exists, we are simply multiplying by zero. This proves the equation.

$$\widehat{\mathbf{E}} \circ \mathbf{L} = \mathbf{T}^C \tag{0}$$

For equation (25), it can easily be seen that if $\widehat{\mathbf{e}}_{i,j} = 1$ for some $i, j$, then no direct link exists between these nodes, and thus no direct flow or alternative flow can exist. This means that the total number of all indirect flows $\mathbf{l}_{i,j}$ will be substitute route flows only. On the other hand, if $\widehat{\mathbf{e}}_{i,j} = 0$, then no substitute flows exist, the product is therefore zero, and the equations hold. This ties in neatly with an equation proved in the previous model, which deals with the OD matrix: $\widehat{\mathbf{E}} \circ \mathbf{D} = \mathbf{T}^C$.

$$\widehat{\mathbf{E}} \circ \widehat{\mathbf{L}} = \mathbf{T}^C \tag{0}$$

$$\widehat{\mathbf{E}} \circ \widehat{\mathbf{D}} = \mathbf{T}^C \tag{0}$$

Equations (0) and (0) are simply variants of the above equations, but using the binarized forms instead. The proofs are similar are will be omitted.

In conclusion, this section showed how multiplying $\mathbf{T}^C$ (or its binarized form) with a matrix that subsumes these matrices (such as $\mathbf{L}$ or $\mathbf{D}$), results in the matrix $\mathbf{T}^C$ itself (or its binarized form, but only if both operands are binarized). In the same vein, multiplying $\widehat{\mathbf{E}}$ with $\mathbf{L}$ or $\mathbf{D}$ produces $\mathbf{T}^C$ (or its binarization, if $\mathbf{L}$ or $\mathbf{D}$ are binarized as well). We note that $\mathbf{L}$ can be decomposed into its two components through $\mathbf{L} = \mathbf{T} + \mathbf{T}^C$. We also note that since a flow from some node to another node can only be either direct or indirect, then $\mathbf{D}$ may be decomposed into $\mathbf{D} = \mathbf{F} + \mathbf{T} + \mathbf{T}^C$ as well.

### C. Alternative route flow-related equations

For alternative route flow, we can derive a similar set of equations to the substitute route flow-related ones.

$$\mathbf{L} \circ \widehat{\mathbf{T}} = \mathbf{T} \tag{0}$$

The proof of equation (0) is similar to that of equation (0) in that it uses the mutual exclusivity of the two types of indirect flows. Because there are only two types of indirect flows, then $\widehat{\mathbf{t}}_{i,j} = 1$ implies that $\mathbf{l}_{i,j}$ pertains solely to the count of alternative route flows, and the equation holds true. (The zero case is trivial.)

$$\widehat{\mathbf{D}} \circ \mathbf{T} = \mathbf{T} \tag{0}$$
$$\widehat{\mathbf{L}} \circ \mathbf{T} = \mathbf{T} \tag{0}$$
$$\widehat{\mathbf{D}} \circ \widehat{\mathbf{T}} = \widehat{\mathbf{T}} \tag{0}$$
$$\widehat{\mathbf{L}} \circ \widehat{\mathbf{T}} = \widehat{\mathbf{T}} \tag{0}$$

Equations (0) and (0) can be proven by showing that $\widehat{\mathbf{d}}_{i,j} = \widehat{\mathbf{l}}_{i,j} = 1$ whenever alternative route flows exist (because the two matrices simply refer to the presence of OD flows and indirect flows, respectively), and 0 if alternative route flows do not exist, thus proving the equation is true for all cases. Equations (0) and (0) simply use the binarized form of **T** and can be proven similarly to the above.

$$\mathbf{A} \circ \mathbf{T} = \mathbf{T} \tag{0}$$
$$\mathbf{A} \circ \widehat{\mathbf{T}} = \widehat{\mathbf{T}} \tag{0}$$

Equations (0) and (0) are interesting in that they now deal with the adjacency matrix. We can prove equation (0) by showing that $\mathbf{t}_{i,j} \geq 1$ implies $\mathbf{a}_{i,j} = 1$, because by definition alternative route flows can only exist in the presence of direct links. This results in a multiplication by 1. The other case is when $\mathbf{t}_{i,j} = 0$, which results in a product of zero. The equation is then proven true, and a similar proof for equation (0) can be created.

As a note, we see more instances of the decomposition of matrices in our Hadamard products. The matrices **L** and **D** both subsume the alternative route flows, and we see similar results as with the substitute route flows: multiplying **T** (or its binarization) with either of the two matrices above or their binarizations results in **T**, but if both operands are binarized, we get the binarization of **T** instead.

### D. Total indirect flow-related equations

There are some additional relationships that are centered on the indirect flow matrix **L** that were not outlined in the previous model.

$$\widehat{\mathbf{D}} \circ \mathbf{L} = \mathbf{L} \tag{0}$$
$$\widehat{\mathbf{D}} \circ \widehat{\mathbf{L}} = \widehat{\mathbf{L}} \tag{0}$$
$$\widehat{\mathbf{L}} \circ \mathbf{L} = \mathbf{L} \tag{0}$$

The proof for these equations can be shown by observing that whenever an indirect flow exists (meaning $\mathbf{l}_{i,j} \geq 0$, or equivalently $\mathbf{l}_{i,j} = 1$), then we are simply multiplying by $\widehat{\mathbf{d}}_{i,j} = \widehat{\mathbf{l}}_{i,j} = 1$, because these binarized matrices simply denote the existence of OD flows or indirect flows, respectively, and in this case the equations holds. In the case where $\mathbf{l}_{i,j} = \widehat{\mathbf{l}}_{i,j} = 0$, we are multiplying by zero, which preserves the equality.

### E. Other equations

In this section, we present a few additional equations that were not covered in the previous model. This section serves to make the extensions to the model as comprehensive as possible. The equations here focus on the flow matrix **F** and the OD matrix **D**.

$$\widehat{\mathbf{F}} \circ \mathbf{F} = \mathbf{F} \tag{0}$$

$$\widehat{\mathbf{D}} \circ \mathbf{D} = \mathbf{D} \tag{0}$$

Equations (0) and (0) are relationships of the same form, and they can be proven true by observing that in these equations we are multiplying elements in **F** (or elements in **D**) by zeros if that same element is zero, and multiplying by one if that element is nonzero. In both cases, the equation holds.

$$\widehat{\mathbf{D}} \circ \mathbf{F} = \mathbf{F} \tag{0}$$
$$\widehat{\mathbf{D}} \circ \widehat{\mathbf{F}} = \widehat{\mathbf{F}} \tag{0}$$

The proof for equation (0) is as follows: if $\mathbf{f}_{i,j} = 0$, then trivially the product is zero. If $\mathbf{f}_{i,j} \geq 0$, then $\widehat{\mathbf{d}}_{i,j} = 1$, and the product is therefore $\widehat{\mathbf{d}}_{i,j} \cdot \mathbf{f}_{i,j} = 1 \cdot \mathbf{f}_{i,j} = \mathbf{f}_{i,j}$ which proves the equation. Equation (0) is proven similarly.

$$\mathbf{A} \circ \widehat{\mathbf{F}} = \widehat{\mathbf{F}} \tag{0}$$

Equation (0) is simply the binarized version of one of the equations in the previous model: $\mathbf{A} \circ \mathbf{F} = \mathbf{F}$. To prove the equation, we simply need to observe that if $\mathbf{a}_{i,j} = 0$, then as a result of not having a direct link, $\mathbf{f}_{i,j} = \widehat{\mathbf{f}}_{i,j} = 0$, and the equality holds; however, when $\mathbf{a}_{i,j} = 1$ then we are simply multiplying by the multiplicative identity, proving the equation true.

### F. Exceptions and wrong equations

In this section, we present a few examples of equations that are are false, to show that deriving these relationships is not as simple as pairing related matrices together.

Let us consider the equation $\widehat{\mathbf{E}} \circ \mathbf{L} = \mathbf{L}$. One may expect this to be correct, considering that $\widehat{\mathbf{E}}$ denotes the possibility of having a substitute route flow, which is a type of indirect route flow. However, this equation is not true for all cases; it is only true for cases where there are no alternative route flows. Consider the following graph in Figure 1.

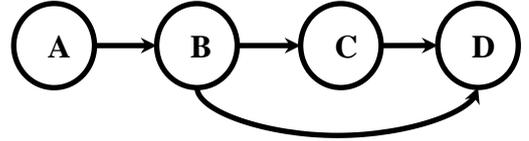

Fig. 1  A directed network graph

If we assume only one trajectory using the network, and this trajectory uses the path A-B-C-D, then it is easy to see that there is an alternative route flow from B to D, that is, $\mathbf{T}_{BD} = 1$. We note that this also means $\mathbf{L}_{BD} = 1$. We now consider $\widehat{\mathbf{E}}$. In this graph, $\widehat{\mathbf{E}}_{BD} = 0$. Therefore, when we multiply, $\widehat{\mathbf{E}}_{BD} \circ \mathbf{L}_{BD} = 0 \neq 1 = \mathbf{L}_{BD}$, thus the equation is not true for all cases.

## VI. CONCLUSION

This paper extended the theory that relates different essential components of traffic analysis, specifically trajectories, (generalized) OD matrices, and flow matrices. We derived new equations in this study and strengthened our understanding of the model to show mutual exclusivity between direct flows and substitute route flows, as well as between alternative route flows and substitute route flows. We showed cases where performing a Hadamard multiplication using two matrices where the first matrix subsumes the second results in the second matrix, essentially decomposing the first matrix into its components and removing everything else but the second. These patterns allow us to easily create and simplify algebraic equations involving these essential network-related matrices to show new relationships.

We can now show a correspondence between the matrices.

TABLE II
GENERAL CORRESPONDENCE FRAMEWORK OF THE NETWORK MATRICES

| Structure | Network Utilization |
|---|---|
| **A** (adjacency matrix) | **F** (flow matrix) |
| **P** (distance matrix) | **D** (OD matrix) |
|  | **L** (indirect flow matrix) |
|  | **T** (alternative route flow matrix) |
| **E** (external matrix) | **$T^C$** (substitute route flow matrix) |

We noted that **E** does not precisely denote the structure for **L**, but for **L** instead: **E** denotes the potential of having substitute route flows. The table shows that we may still need to find structural matrices for **L** and **T**, and definitions and algorithms for solving their values.

Further work that is being done investigates the existence of cycles within the trajectory inputs, as well as exploring additional structural matrices that correspond to the indirect flow matrix and the alternative route flow matrix.

ACKNOWLEDGMENT

This research was funded by the Commission on Higher Education (CHED) Philippine Higher Education Research Network (PHERNet).